\documentstyle[12pt]{article}

\normalbaselines 
\newcommand{\Ir}{Z\!\!\!Z}
\newcommand{\beq}{\begin{equation}}
\newcommand{\eeq}{\end{equation}}
\newcommand{\Ibb}[1]{ {\rm I\ifmmode\mkern 
            -3.6mu\else\kern -.2em\fi#1}}
\newcommand{\ibb}[1]{\leavevmode\hbox{\kern.3em\vrule
     height 1.2ex depth -.3ex width .2pt\kern-.3em\rm#1}}
\newcommand{\Nl}{{\Ibb N}}
                                          
\newcommand{\lapprox}{\stackrel{\textstyle<}{\sim}}

\newcommand{\bi}{b_{ik}}

\begin{document} 
\begin{center}
\vspace*{1.0cm}

{\LARGE{\bf Emergence of Space-Time on the Planck Scale described as
    an Unfolding Phase Transition within the Scheme of Dynamical
    Cellular Networks and Random Graphs}} 

\vskip 1.5cm

{\large {\bf Manfred Requardt }} 

\vskip 0.5 cm 

Institut f\"ur Theoretische Physik \\ 
Universit\"at G\"ottingen \\ 
Bunsenstrasse 9 \\ 
37073 G\"ottingen \quad Germany

\end{center}

\vspace{1 cm}

\begin{abstract}
As in an earlier paper we start from the hypothesis that
physics on the Planck scale should be described by means of concepts
taken from {\it
  discrete mathematics}. This goal is realized by developing a scheme
being based on the dynamical evolution of a particular class of {\it
  cellular networks} being capable of performing an {\it
  unfolding phase transition} from a (presumed) chaotic initial phase towards a
new phase which acts as an {\it attractor} in total phase space and
which carries a fine or super structure which is identified as the discrete
substratum underlying ordinary continuous space-time (or rather, the
physical vacuum). Among other things we analyze the internal structure
of certain particular subclusters of nodes/bonds (maximal connected
subsimplices, $mss$) which are the fundamental building blocks of this
new phase and which are conjectured
to correspond to the {\it
  physical points} of ordinary space-time. Their mutual entanglement
generates a certain
near- and far-order, viz. a causal structure within the network which
is again set into relation with the topological/metrical and
causal/geometrical structure of continuous space-time. The
mathematical techniques to be employed consist mainly of a blend of a fair amount of {\it stochastic mathematics} with several
relatively advanced topics of discrete mathematics like the {\it
  theory of random graphs} or {\it combinatorial graph theory}. Our
working philosophy is it to create a scenario in which it becomes
possible to identify both gravity and quantum theory as
the two dominant but derived(!) aspects of an underlying discrete and
more primordial
theory (dynamical cellular network) on a much coarser level of
resolution, viz. continuous space-time.

\end{abstract} \newpage
\section{Introduction}
\noindent In a previous paper (\cite{1}), starting from the hypothesis
that both physics and mathematics are discrete on the Planck scale, we
developed a certain framework in form of a class of '{\it cellular
  network models}' consisting of cells (nodes) interacting with each
other via bonds according to a certain '{\it local law}' which governs
their evolution.

Our personal philosophy in this endeavor is the following: If one
wants to go beyond the metric continuum (a possibility which was
already contemplated by B.Riemann in \cite{2}), an entirely different
and in some respects perhaps even new arsenal of physical/mathematical
concepts and calculational tools is called for respectively has to be
invented. (Ideas in a spirit similar to ours can e.g. be found in the
work of R.Sorkin et al; \cite{7} or C.Isham; \cite{11}. A slightly
different line of thought is persued in e.g. \cite{10}; see further
remarks at the end of section 2). Among other things one has to be extremely careful not to reimport
notions and ideas through the backdoor which carry openly or
implicitly a meaning or connotation having its origin in continuum
concepts and, more generally, to obey Occam's principle, i.e. stick
to as few as possible clear and natural hypotheses.

A case in point is the concept of {\it dimension} which appears to be, at
least traditionally, a typical continuum concept. As ordinary physics
can successfully be described by employing a framework which has as
one of its cornerstones the postulate of space-time as a {\it metric
  continuum}, a serious candidate for an underlying discrete and more
primordial substratum shall have the capability to generate a concept
like dimension as an '{\it emergent}' collective quality without(!)
having it already among its elementary buildingblocks in one or the
other disguise.

We showed in \cite{1} that our class of discrete model theories do
have indeed a rich enough structure to accomplish, among other
things, this goal. More specifically, scrutinizing what are in fact
the crucial ingredients of something like dimension from the point of
view of physics, in particular dynamics and interaction in physical
systems, we managed to develop a notion of '{\it generalized
  dimension}' also in discrete and quite irregular systems and which
mirrors both a characteristic and '{\it intrinsic}' property of the
discrete network models under discussion by measuring its '{\it
  connectivity}' and, on the other side, does not resort to some sort
of underlying dimension of a possible '{\it embedding space}'. In this
sense it is in exactly the same way an '{\it intrinsic}' concept as,
say, the concept of '{\it intrinsic curvature}' for
manifolds. Furthermore it is indeed a true generalisation as it
coincides in the regular and ordinary situations with the usual
definition of dimension.\\[0.5cm]
Remark: One can introduce of course other concepts over discrete
structures which carry a certain dimensional flavor but which have
their basis not so much in physics and dynamics but rather in the
algebraic topology of, say, '{\it simplicial complexes}'
(cf. e.g. \cite{8}). To some extent such ideas were also persued in
\cite{1} but we have the (subjective) impression that the '{\it
  connectivity-dimension}' of the network is the really crucial
property from the physical point of view (in particular when something
like '{\it unfolding phase transitions}' and other large scale phenomena become
relevant). On the other hand these different concepts may neatly
complement each other in encoding appropriately various facets of
'{\it complex systems}'.\\[0.5cm]
In \cite{1}, besides introducing our network concept, possible
dynamics on it and defining the concept of connectivity dimension, our
main concern was the development of a kind of '{\it discrete
  analysis}' and the clarification of its relation to various other modern
approaches which follow different lines of reasoning but seem,
nevertheless, to be inspired by the same philosophy, i.e. the complex
of ideas and methods called '{\it non-commutative geometry}'. On the
other side, the possible dynamical processes in our network models,
which are expected to lead, among other things, to the emergence of
something like '{\it space-time}' and continuum physics in general
were only briefly sketched.

The main purpose of this paper is a more detailed investigation of
this '{\it unfolding}' of '{\it phase transition type}' and the
development of the necessary mathematical, mostly
topological/geometric, tools and concepts which will allow us to cast
vague and qualitative reasoning into a more rigorous and precise
framework.\\[0.5cm]
Remark: As to references to the field in general cf. e.g. \cite{1} and
further references given in the papers mentioned there. A deserving
bibliography of papers dealing with more or less non-orthodox ideas
about space-time and related concepts is \cite{3}. Various
further mathematical sources we have found useful for our approach
will be cited below.\\[0.5cm]
A further clarifying remark (a kind of ''petitio principii'')
concerning the (mathematical) rigor of the type of reasoning we are
going to present in the following seems to be in order. In some sense
one should regard it rather as a blending of some ''educated
speculation'' with various pieces of more rigorous mathematics. The
reasons for this are, as we think, almost inescapable. For one, we
have to set up and fix both the complete abstract framework and
conceptual tools and then to infer practical, observable consequences
for ordinary space-time physics more or less at the same time in order
to motivate our primordial assumptions. This is already difficult
enough as we have to close a gap between Planck-scale physics and,
say, ordinary quantum physics which extends over many orders of
magnitude. On the other side we consider this to be of tantamount
importance since it is, given the complete absence of reliable
experimental data, the only selection principle allowing to judge the
scientific value of the various assumptions in our model building
process.

For another, many of the mathematical disciplines to be employed in
the following are either almost in their infancies (like e.g. the
correct statistical treatment of extremely complex and densely
entangled dynamical systems, consisting of an
extraordinarily large number of degrees of freedom, typical cases in
point being our cellular networks), or do not belong to the usual
armory of mathematical physics (as e.g. various branches of discrete
mathematics) and have to be developed to a certain extent almost from
scratch, at least as applications to our field are concerned.

Therefore our general strategy will be as follows: we will prefer to
develop in this paper what we think is the generic scenario in broad
outline up to the point that we can make transparent how several
(partly long standing and puzzling) problems of ordinary space-time
physics might be successfully approached within our much wider
framework with, among other things, its deeper conception of the
nature and role of space-time and the corrresponding fine structure of
the physical vacuum.

There remain quite a number of technical details to be filled in,
sometimes even veritable subtheories, which may have some interest of
their own. This shall in part be postponed to subsequent work in case
it would blow up the paper to much and after the general scheme has
proved its value. For the time being we will content ourselves with
inserting some preliminary remarks at appropriate places in the
following sections which are to serve as hints as to where more detailed
investigations seem to be worthwhile. In our view the central section
of this paper (both with respect to applied mathematical techniques
and physical content) is section four. 

Concluding the introduction we would like to add the remark that after
having nearly completed this paper we stumbled by chance over an
exceptionally beautiful (but, as is probably the case with the whole
discipline, perhaps not so widely known) book by Bollobas
(\cite{Bollo2}; unfortunately it is even out of print at the moment)
about '{\it random graphs}', a field originally founded by Erdoes and
Renyi in the late fifties. By brousing through the book we realized
that the results, being presented there, may confirm, if appropriately
translated, to a large extent the soundness of several of our
speculations about the assumed generic behavior of our unfolding
network. In section 4 of our paper we briefly mention some results
from this book; we plan however to embark on a closer inspection of
the relations between these two fields in subsequent work (note that
the theory of random graphs was founded to tackle some deep problems
in graph theory proper, not so much for its probabilistic aspects or even
possible applications in physics).
\section{The Class of Dynamical Cellular Networks}
The idea to base '{\it discrete dynamics}' on the Planck scale on
something like a cellular network was partly inspired by the role
which its close relatives, the cellular automata, are playing in the
science of complexity (see the references given in e.g. \cite{1}) and
the general philosophy to achieve complex, emergent behavior and
pattern generation by means of surprisingly simple looking microscopic
'{\it local laws}'.

On the other side, a cellular automaton proper appeared to be a far too
regular and rigid array in our view, in particular in the context of
possible formation of something like space-time as an ordered
superstructure being embedded in or floating above a less organized
and probably almost chaotic substratum. More specifically, the
organisation of the network as such should be pronouncedly dynamical,
most notably its kind of wiring. Furthermore, we maintain that really
fundamental dynamical laws must have a quite specific structure, the
basic ingredient being the mutual dynamical coupling of more or less
two distinct classes of entities, in this context the one being of a local
character
(nodes or sites), the other of a pronouncedly ''geometrical'' quality
(bonds or links); the state of the one class being governed by the
previous state of the other class. Typical cases in point are
e.g. general relativity and gauge theory. In contrast to our network
models the bonds in a cellular automaton are typically both rigid and
regular and not(!) dynamical so that a '{\it backreaction}' of the
node states (''matterfields'') on the bond states
(''geometry/curvature'') is supressed.

To achieve this we assumed our system to be made up of '{\it cells}'
and elementary interactions mediating among these cells, both of which
are taken to be dynamical variables. This qualitative picture is then
condensed into the concept of a '{\it network}' consisting of '{\it
  nodes}' and '{\it bonds}' (synonyma being sites, vertices or links,
edges respectively) with each bond connecting two {\it different}
nodes.

At each node $n_i$ ($i$ running through a certain index set) the
corresponding cell can be in a specific internal state $s_i \in {\cal
  S}$ (${\cal S}$ typically being some countable or finite set not
further specified at the moment). Correspondingly the bonds
$\{b_{ik}\}$, $b_{ik}$ the bond between node $n_i$ and node $n_k$,
carry a dynamical '{\it valuation}' $\{J_{ik} \in {\cal J}\}$. A
dynamical '{\it local (global) law}' $ll$, ($LL$) is then introduced
via the general evolution equation given below.\\[0.5cm]
Remark: For the time being we employ a synchroneous '{\it clock
  time}'\,$t$, proceeding in discrete elementary steps $\tau$,
i.e. $t=n\cdot\tau$. That means, the whole system is updated in
discrete steps. This clock time should {\it not}(!) be confused with
socalled '{\it physical time}' which is expected to arise as an
emergent collective concept on a possibly much coarser scale. In any
case we conjecture that 'local' physical time will turn out to be some
kind of '{\it order parameter field}' coming into existence via the
dynamical interaction of many nodes/bonds within the local grains
under discussion (see below). The choice of an external overall clock
time is made mainly for technical convenience in order to keep already
quite complicated matters reasonably simple and can be modified if
necessary.\vspace{0.5cm}

To sum up, we make the following definitions:\\[0.5cm]
{\bf 2.1 Definition (Cellular Network)}: In the following we will deal with the class of systems defined below:\\
i) "Geometrically" they are {\it graphs}, i.e. they consist of nodes
\{$n_i$\} and bonds \{$\bi$\} where pictorially the bond $\bi$
connects the nodes $n_i$ and $n_k$ with $n_i\neq n_k$ implied (there
are graphs where this is not so), furthermore, to each pair of nodes
there exists at most one bond connecting them. In other words the
graph is 'simple' (schlicht). There is an intimate relationship
between the theory of graphs and the algebra of relations on sets. In
this latter context one would call a simple graph a set carrying a
homogeneous non-reflexive, (a)symmetric relation.

The graph is assumed to be {\it connected}, i.e. two arbitrary nodes
can be connected by a sequence of consecutive bonds, and {\it
regular}, that is it looks locally the same everywhere.
Mathematically this means that the number of bonds being incident
with a given node is the same over the graph ({\it 'degree' of a node}).
We call the nodes which can be reached from a given node by making
one step the {\it 1-order-neighborhood} ${\cal U}_1$ and by not more
than n steps ${\cal U}_n$.\\
ii) On the graph we implant a class of dynamics in the following way:\\[0.5cm]
{\bf 2.2 Definition(Dynamics)}: As for a cellular automaton each node
$n_i$ can be in a number of internal states $s_i\in \cal S$. Each
bond $\bi$ carries a corresponding bond state $J_{ik}\in\cal J$. Then
we assume:
\begin{equation} s_i(t+\tau)=ll_s(\{s'_k(t)\},\{J'_{kl}(t)\})
\end{equation}
\begin{equation} J_{ik}(t+\tau)=ll_J(\{s'_l(t)\},\{J'_{lm}(t)\})
\end{equation}
\begin{equation}
(\underline{S},\underline{J})(t+\tau)=LL((\underline{S},\underline{J}(
t))
\end{equation}
where $ll_s$, $ll_J$ are two mappings (being the same all over the
graph) from the state space of a local
neighborhood of a given fixed node or bond to $\cal S, J$, yielding the
updated values of $s_i$ and $\bi$. $s'_l$ and $J'_{lm}$ denote the
internal states of the nodes and bonds of this neighborhood under
discussion, $\underline{S},\underline{J},LL$ the corresponding global
extended states and law.\\[0.5cm]
Remark: The theory of graphs is developed in e.g. \cite{4,5}. As
to the connections to the algebra of relations see also
\cite{6}. There exist a lot of further concepts in '{\it graph theory}' and '{\it
  discrete mathematics}' in general which are
useful in our context, some of which (together with some further
relevant references )will be introduced below, especially in section 4.\\[0.5cm]
{\bf 2.3 Lemma}: The kind of discrete analysis we developed in
\cite{1} made it necessary to give the bonds an '{\it orientation}'
(which is extremely natural anyhow). Consistency then requires an
analogous relation to hold for the interactions, i.e:
\begin{equation} b_{ik}=-b_{ki} \quad \mbox{and hence} \quad
  J_{ik}=-J_{ki} \end{equation} 
\vspace{0.3cm}
It is now our aim to attempt to motivate a kind of unfolding and
pattern creation in such a network, which, starting from a certain
initial state (or better: phase), is supposed to lead in the coarse of
a dynamical process -- among other things -- to the emergence of
something we are experiencing on a coarser scale of resolution as
'{\it space-time}'. In order to achieve this we have in a first step
to model a kind of concrete '{\it critical}' dynamics which, on the
one hand, fits in the general scheme of Definition 2.2. On the other
hand it turns out to be a subtle task to appropriately implement the
crucial ingredients which are to catalyze the unfolding process.

We presented several types of local laws in \cite{1}. We in fact
studied quite a few other possible dynamical laws, the corresponding
scenarios and mappings between the mathematical entities occurring in
these scenarios and the concepts  carrying a certain physical meaning
on a possibly much coarser scale of resolution like e.g. space-time as
such, matter, (quantum)fields, gravitation etc. The main difficulty was to implement the unfolding process of our '{\it
  network universe}' in a natural way and understand as what kind of
  '{\it order parameter manifold}' space-time was going to emerge in
  this evolving and quite chaotic background. In this selection
  or identification process an important role was played by '{\it
  Occam's principle}', i.e. to infer seemingly complicated phenomena
  from as simple as possible and, as we think, uncontorted microscopic
  laws. A crucial ingredient is the possible switching-on or -off of
  bonds $b_{ik}$, more specifically, of the corresponding elementary
  interactions $J_{ik}$ in a '{\it hystheresis like}' manner (in
  catastrophy theory called a '{\it fold}').
 
These criteria can
  probably be fulfilled by a whole class of dynamical network laws, or
  expressed more carefully: there may exist a whole class of possible
  candidates, the behavior of which could be checked by simulation on a
  computer, from which we are presenting in the following  the probably most simple ones. At the
  moment implementations of various possible network models on a computer
  are under the way. These investigations are quite time consuming for
  various reasons (complex behavior is difficult to study anyhow;
  cf. e.g. reference \cite{17}), viz. it may turn out that the simple
  laws we are going to describe below have to be replaced in the end by another law
  from the same class in case they e.g. do not behave chaotically enough. These
  numerical investigations will be presented elsewhere.

The naive picture encoded in the following law is that a, for the time being, not further specified
substance consisting of elementary quanta $q$ is transported through
the network according to the following law:\\[0.5cm]
{\bf 2.4 Definition of the 'Unfolding' Local Law}: The local state
space at each site $n_i$ is assumed to be either 
\begin{equation} {\cal S} :=\{n\cdot q;0\leq n\leq
  N,n\in\Ir\}\quad\mbox{or}\quad q\cdot\Ir\end{equation}
The bonds $b_{ik}$ can carry the '{\it valuation}'
\begin{equation} J_{ik}\in \{-1,0,+1\} \end{equation}
Remark: As to the possible allowed range of the variables $s_i$ some
additional remarks are in order which will be given below.\\[0.5cm]
Then the first half of the dynamical local law reads in the case
${\cal S}=q\cdot\Ir$:\\
{\bf A})\begin{equation}s_i(t+\tau)-s_i(t)=-q\cdot\sum_k
  J_{ik}(t)\end{equation} 
the sum extending over all the bonds being incident with the node
$n_i$.\\
The second half, describing the backreaction of the node states on the
bond states, is slightly more complicated:\\
{\bf
  B})\begin{equation}\mbox{1)}J_{ik}(t+\tau)=-sgn(s_k(t)-s_i(t))\;\mbox{if}\;J_{ik}(t)\neq 0\;\mbox{and}\end{equation}
\begin{equation}\delta s_{ik}:=|s_k(t)-s_i(t)|\geq\lambda_1\;\mbox{or}\;J_{ik}=0\;\mbox{and}\;\delta s_{ik}(t)>\lambda_2\end{equation}
with\begin{equation}0<\lambda_1<\lambda_2\quad\mbox{and}\quad
  I_{\lambda}:=[\lambda_1,\lambda_2]\end{equation}
denoting the '{\it hystheresis interval}'.
\begin{equation}\mbox{2)}J_{ik}(t+\tau)=0\quad\mbox{if}\quad\delta
  s_{ik}<\lambda_1\end{equation}
To say it in words: Below $\lambda_1$ bonds are temporarily
annihilated, above $\lambda_2$ they are turned on again, and in between
the bonds may only switch between $+1$ and $-1$, which is essentially a
switch in orientation or local direction of transport which can be
seen from the first half of the law.\\[0.5cm]
{\bf C)} The roles of $\lambda_1,\;\lambda_2$ are interchanged,
i.e. bonds $b_{ik}$ are switched off for $\delta s_{ik}>\lambda_2$ and
switched on again for $\delta s_{ik}<\lambda_1$.\\[0.5cm]
Remark: Both possible laws are presently carefully studied on the
computer. As the simulations will take some time, the results will be
published elsewhere. What one can however already say is that one of
the many fascinating observations is the ability of systems like these
to self-organize themselves and to find a variety of different
attractors, i.e. to typically occupy a relatively small region of the
potentially accessible huge phase space. Furthermore type {\bf C)}
leads to a totally different behavior as compared with type {\bf B)}.\vspace{0.5cm}

${\cal S}:=q\cdot\Ir$ is probably appropriate if one does not want to
bother about boundary conditions and should be considered as the
idealisation of the scenario where typical local fluctuations of $s$,
i.e.
\begin{equation} \Delta s:=<(s-<s>)^2>^{1/2} \end{equation}
the average taken over some suitable time sequence and/or suitably
large spacial array of nodes, are sufficiently far away from the
possible internal boundaries of the concrete system with a finite
local state space, e.g.
\begin{equation}s\in \{n\cdot q,\quad 0\leq n\leq N\gg 1\}\end{equation}
One could hence regard the former system as a certain limit for $N\to
\infty$ and shifting at the same time the reference point from, say,
$N/2$ to zero.\vspace{0.5cm}

On the other side, for finite $N$ one can introduce boundary
conditions like the following ones, arguing that they do not change
the qualitative behavior provided $N$ is much larger than the generic
local fluctuations about some average value $<s>$ with
\begin{equation}0\ll<s>\ll N\quad\mbox{,say, and}\quad <s>={\cal O}(N/2)
\end{equation}
{\bf 2.4 A')}:
\begin{equation}s_i(t+\tau)=s_i(t)\quad\mbox{if}\quad -q\cdot\sum_k
  J_{ik}(t)>q\cdot N-s_i(t)\quad\mbox{or}\quad
  q\cdot\sum_kJ_{ik}(t)>s_i\end{equation}
i.e. there is no transport if it would transcend the maximal/minimal
  '{\it charge}' of the node, $Q=q\cdot N$ or $0$.\\[0.5cm]
{\bf A'')}: Perhaps more natural are '{\it periodic boundary
  conditions}', i.e. the above equation has to be understood modulo $N$.\vspace{0.5cm}

We want to conclude this section with a couple of additional remarks
as to this particular type of local law:\\[0.5cm]
Remarks: i) The above dynamical law shows that under certain
circumstances '{\it elementary interactions}' $J_{ik}$ may become zero
for some lapse of time. It may even be possible that a substantial
fraction becomes '{\it locked in}' at the value zero for a rather long
time. This will then have some practical consequences for the
(partial) representation of our network as a graph. On the one side
one may consider the graph, in particular its wiring, to be given as a
fixed static underlying substratum, i.e. its bonds $b_{ik}$ ''being there'' as
elements of the graph even if a substantial fraction of the
corresponding elementary interactions $J_{ik}$ is zero temporarily or
even locked in at zero.

On the other side, if one wants to deal with e.g. phase transition
like '{\it topological/geometrical}' changes of the whole wiring of
the network it may be advantageous to change the point of view a
little bit and regard not only the bond valuations $J_{ik}$ as
dynamical variables but rather the underlying graph as such, more specifically,its bonds $b_{ik}$, i.e. allow them to be annihilated or
created. As a consequence the underlying graph will change its shape
in the course of clock time which may perhaps be a more fruitful mode
of representation. This latter point of view will be adopted in the
following where it seems to be appropriate.\\
ii) There is a certain (faint) resemblance to non-linear electrical
networks. One can in fact, if one likes so, consider the $S$-field as
a charge distribution (or potential-field) with $s_i-s_k$ as kind of
voltage difference between neighboring nodes. Part A) of the dynamical
law reflects then the usual charge conservation. The interpretation of
B) is more complicated. One could e.g. regard it as a non-linear
dependence of the resistances of links on the applied voltages. In any
case, we are confident that it be possible to realize or emulate our
system as some kind of non-linear network, thus producing possibly
effects which resemble the unfolding of our universe in the
laboratory!\\
iii) We would like to emphasize that our dynamical law is a genuine
cellular network law and not(!) some Lagrangian field theory in
disguise (which is frequently the case in other approaches). In our
view it is not at all selfevident that it be advisable to model
fundamental laws on the Planck scale according to one or the other
kind of an (at best) '{\it effective quantum field theory}' which
lives many scales above the Planck regime.

On the other side, we are quite confident that these known types of
field theories will emerge as effective theories after some
appropriate kind of
'{\it renormalisation transformation}' on a much coarser level,
describing the interaction of patterns (e.g. fields) which are
themselves rather '{\it collective excitations}'. In this sense our
approach is very much in the spirit of the philosophy of t'Hooft
(cf. \cite{12}).\\
iv) Our personal philosophy is ''iconoclastic'' in the sense described
in the beautiful reviews about quantum gravity by Isham (see
e.g. \cite{9}) in sofar as we want to show that and how both quantum
(field) theory and gravitation emerge as two different but related
aspects of one and the same underlying and more primordial theory of
the kind introduced above. The first steps of this endeavor will be
undertaken in this paper, which is mainly concerned with the analysis
of the kind of embedding of a certain '{\it order parameter manifold}'
$ST$ (space-time) in the background space $QX$ (the cellular
network). Most importantly we are concerned with the internal
structure of certain subclusters of nodes/bonds which are on the
''continuous'' (coarse grained) level ''experienced'' as physical
points and their mutual causal relations.

In this context a lot of topological analysis has to be carried out
which is performed in a similar spirit as the work of Sorkin et
al. (\cite{7}) or Isham (\cite{11}) at least as far as the general
philosophy is concerned. The technical tools and concepts being
employed may however be different. In our case it is mainly a blend of
various fields of '{\it discrete mathematics}' as e.g. graph theory,
finite geometry, calculus of relations and the like with arguments
boroughed from statistical physics/mathematics and the science of
complexity. Some of the concepts developed below may even be new but
this is difficult to decide from the widely scattered literature, a
substantial fraction of which may have escaped our notice
up to now.  
\section{The Primordial (Chaotic) Network Phase $QX_0$}
From the structure of the dynamical law we have elaborated in section
2 we can infer that our dynamical network, which we henceforth denote
by $QX$ ('{\it quantum space}'; at the moment this should however only
be considered as a metaphor), is at each fixed clock time $t$ a
certain (dynamic) graph $G(t)$ if we do not take into account the
details of the status of the fields of node and bond states but
concentrate solely on its purely geometric content. The dynamics of
$G(t)$ consists of the possible switching off or on of some bonds in
the time step from $t$ to $t+\tau$.\\[0.5cm]
{\bf 3.1 Definition}: The full dynamical network (with its
distribution of bond and node states being included) we call $QX$ or
$QX(t)$. If we want to concentrate on its purely geometric content we
view it as a graph $G$ or $G(t)$.\vspace{0.5cm}

Our dynamical law is up to now '{\it deterministic}'. Therefore we
should specify the kind of '{\it initial state}' (or rather a generic
group of initial states) from which our network is supposed to
evolve. As was the case with the class of admissible dynamical laws,
we experimented with quite a few types of possible initial states,
looking to what kind of scenarios they would probably lead. It is our
impression that both the most natural (having Occam's principle in
mind) and most promising assumption is to start from a maximally
connected graph, a socalled '{\it complete graph}' or '{\it simplex}',
i.e. each two nodes $n_i,n_k$ are connected by a bond $b_{ik}=-b_{ki}$
or (see below) from a class of graphs which are '{\it almost
  complete}'.\\[0.5cm]
{\bf 3.2 Assumption}: We assume that our '{\it initial phase} $QX_0$
consists generically (for reasons explained below) of '{\it almost
  complete graphs}', i.e. the '{\it average number}' of active bonds
($J_{ik}\neq 0$) is almost maximal (see the following
remarks).\vspace{0.5cm}

Some remarks are here in order: One knows already from the analysis of
cellular automata that these systems behave typically in an extremely
complicated manner and that for exactly the same reasons as in
e.g. statistical mechanics it makes frequently only sense to study
'{\it generic}' properties and behavior, mostly with the help of
statistical means (cf. e.g. the references on cellular automata in
\cite{1}, in particular the paper by Wolfram; \cite{13}).

In other words, it seems more appropriate to speak of different '{\it
  phases}' instead of individual states. Our dynamical law shows that,
  typically, a certain percentage of bonds are switched on or off at
  each clock time step or during a certain interval. That means, it is
  more natural to assume that $QX$ was ''initially'' or better:
  remained for possibly quite some time in a phase with two arbitrary
  nodes being connected on average with a probability '{\it near
  one}'. If we assume that the number of nodes in our network is a very
  large but finite number $\Lambda$, a simplex has
  $\Lambda(\Lambda-1)/2$ bonds. This means:\\[0.5cm]
{\bf 3.3 Assumption (Statistical Version)}: We assume that our network
  evolves from an initial phase $QX_0$ with the average number of
  bonds (taken e.g. over a suitable clock time interval) being
  ''near'' $\Lambda(\Lambda-1)/2$.\vspace{0.5cm}

Henceforth our arguments will frequently (for obvious reasons) carry a
markedly statistical flavor. Therefore we will introduce a couple of
corresponding concepts and notations.\\[0.5cm]
{\bf 3.4 Definition}: i) We denote the set of vertices (nodes) and
edges (bonds) by $V$, $E$ respectively (we conform here to the usual
notations of graph theory; note that in our case - simple graphs - $E$
can be viewed as a certain subset of $V\times V$ and represents a
homogeneous, non-reflexive, symmetric relation), their cardinalities
by $|V|,|E|$. The degree of a node $n$, i.e. the number of bonds being
{\it incident} with it, is $deg(n)$. The graph is called {\it regular}
if $deg(n)$ is constant over the whole graph.\\
ii) As in our case $deg(n)$ or $|E|$ are time dependent (if we adopt
the point of view developed in Remark i) at the end of the previous
section, i.e. consider the underlying graph as such as a dynamical
object), it makes sense to build the corresponding statistical
concepts:
\begin{equation}<deg>_s,\, <deg(\circ)>_t,\, <deg>_{st},\,
  <E>_t\end{equation}
are the respective averages of $deg$, $E$ with $<\circ>_s$ the
''spacial'' average over the graph at fixed clock time $t$,
$<\circ>_t$ some temporal average (e.g. at a fixed node),
$<\circ>_{st}$ a space-time average. The time average is assumed to be
taken over an appropriate time interval, the length of which depends
on the specific context under discussion (e.g. the particular '{\it
  phase}', various correlation lenghts etc.).\\
Our philosophy concerning the possible time dependence is, for the
time being, that a bond $b_{ik}$ is contributing in, say, $deg$ or
$|E|$ at clock time $t$ if $J_{ik} \neq 0$. 
\vspace{0.5cm}

Evidently there exist certain relations between these notions,
e.g:\\[0.5cm]
{\bf 3.5 Observation}: With 
\begin{equation}<deg>_s:=\sum_V deg(n)/|V|\;\mbox{and}\;\sum_V
  deg(n)/2=|E|\end{equation}
we have
\begin{equation}<deg>_s\cdot |V|/2=|E|\end{equation}
and
\begin{equation}<deg>_{st}\cdot |V|/2=<E>_t\end{equation}
Remark: In \cite{1} we gave each bond an orientation
s.t. $b_{ik}=-b_{ki}$. We are however counting in Observation 3.5
these two differently oriented bonds as one and the same bond (as
usual), hence the factor $1/2$. In any case, this does not make a big
difference as one can always associate an '{\it undirected}' graph
with a '{\it bidirected}' graph.\\[0.5cm]
{\bf 3.6 Conclusion}: With the help of what we have said above,
Assumption 3.3 can now be framed this way:\\
$<deg>_s$ and/or $<deg>_{st}$ remain in a small neighborhood of
$(|V|-1)$ in the '{\it initial phase}' $QX_0$ for a possibly long
time, where ''small'' has to be understood pragmatically, depending on
the details of the model and the context.\vspace{0.5cm}

Our network (graph) $QX$ ($G$) carries a natural metric which makes it into
a '{\it metric space}'. This was already employed in \cite{1}, to
which the reader is referred for further details. Assuming that $G$ is
connected (i.e. every two nodes can be connected by a bond sequence)
there exists a '{\it path}' of minimal '{\it length}' (i.e. number of
bonds). Then we have:
\begin{equation}d(n_i,n_k):=min(\mbox{length of bond sequences, connecting}\;n_i,n_k)\end{equation}
defines a metric on $G$.

With this metric $G$ becomes a discrete topological space (even a
'{\it Hausdorff space}') with a natural neighborhood structure of a
given node, i.e:
\begin{equation}U_m(n_0)=\{nodes;\;d(n_0,n)\leq m\}\end{equation}
The topology as such is however relatively uninteresting since in a
discrete, finite Hausdorff space each one-element set is necessarily
both open and closed. The simple proof runs as follows:
\begin{equation}n_0\neq n_i\;\mbox{implies}\; \exists\;\mbox{open set}\; O_i,n_0\in O_i,n_i\notin O_i\end{equation}
$G$ is finite hence $\cap O_i$ is a finite intersection of open sets
and hence again open. By construction $\cap O_i$ contains no points
other than $n_0$, that is $n_0$ is open. On the other side, with $\cup
n_i$ open, $n_0=G-\cup n_i$ is also closed.

But despite of this, '{\it finitary}' topological spaces are not(!)
automatically trivial as can be seen from the beautiful analysis
performed by Sorkin \cite{7}. However they are typically only socalled
$T_0$-spaces
 (as to these topological notions cf. any good
  textbook on topology as e.g. \cite{14}).

At this place we want to make a proviso. The existence of a
(physically) natural metric on $G$ and a corresponding natural
neighborhood structure neatly implementing the physically important
concepts of ''near by'' and ''far away'' is perhaps much more
important than the existence of a (mathematically) non-trivial
topological structure. It may well be that on such genuinely discrete
sets like networks or graphs natural(!) topological/geometric concepts
should rather be adapted to the real discreteness of the substratum
and perhaps not so much to an abstract axiomatic system defining a
topology. This point of view will be more fully developed in the
following section, to which we postpone the details of the (partly
intricate) topological/geometric analysis.

In the rest of this section, however, we will try to sketch the
expected qualitative behavior of our network $QX$ by employing both statistical
concepts and the concrete local law introduced above. Of particular
interest in this respect are the initial phase $QX_0$ and the scenario
when it starts to leave this  (possibly '{\it metastable}') phase due
to some sort of phase transition. The analysis is, for the time being,
necessarily of a qualitative/statistical nature as only generic
aspects should matter, especially as we are at the moment mainly
interested in deriving observable consequences on much coarser scales
which should not depend to much on microscopic details. For another -
as is the case in statistical mechanics - it would be technically
impossible (at least at the moment) to justify too detailed
assumptions and draw precise conclusions from them.

Let us briefly recall the assumptions made about the phase $QX_0$:
\begin{equation}\mbox{i)}\,|V|=\Lambda\,\mbox{extremely
    large,}\;|E|\approx \Lambda(\Lambda-1)/2\,\to\,<deg>\approx
    \Lambda-1
\end{equation}
i.e. $QX_0$ is almost a simplex or complete graph.\\
ii) The local state space at node $n_i$ consists of multiples of an
elementary quantum $q$, i.e:
\begin{equation}s\in \{q\cdot
  n\},\;n\in\Nl\;\mbox{or}\;\Ir\end{equation}
ranging from $0$ to an appropriately large number $N$ or from $-N$ to
$+N$ (cf. section 2). If we do not want to bother about boundary
conditions we assume the accessible local state space to be the full
$q\cdot\Ir$.\\
iii) The bonds $b_{ik}$ carry the valuation $\{-1,0,+1\}$ modelling the
elementary interactions $J_{ik}$.\\
iv) At each clock time step $\tau$ either an elementary quantum $q$ is
transported via $b_{ik}$ depending on the sign of $J_{ik}$ or the bond
$b_{ik}$ is ''dead'' if $J_{ik}=0$. The bond states $J_{ik}$ at clock
time $t+\tau$ are dynamically coupled with the ''charge difference''
of the incident site states $s_i-s_k$ at clock time $t$.\\
v) Of crucial importance for pattern formation is the built-in
''hysteresis law'' and the hysteresis interval
\begin{equation}[\lambda_1,\lambda_2]\;,\;\lambda_1\;,\;\lambda_2\quad\mbox{being
    the lower and upper critical parameter}\end{equation}
For\begin{equation}\delta
  s_{ik}<\lambda_1\;\mbox{or}\;>\lambda_2\end{equation}
the corresponding bond becomes extinct in the next step or becomes
alive again if it had been extinct before, respectively the other way
around in case of law {\bf 2.4 C)}. In the following we discuss law
{\bf 2.4 B)}. The qualitative reasoning in case of {\bf C)} would
follow similar lines.\vspace{0.5cm}

If the local fluctuations $\delta s_{ik}$ are sufficiently large on
average, i.e:
\begin{equation}\Delta s_{ik}:=<(s_i-s_k)^2>^{1/2}\,>\lambda_2\end{equation}
the corresponding phase is more or less stationary (a slightly more
detailed analysis can be found below). Inspecting the
situation assumed to prevail in the phase $QX_0$ with
$<deg>\approx\Lambda-1$ extremely large, we may infer that also the
average fluctuation of the local charge $s_i$ at an arbitrary node
$n_i$ is typically very large at each clock time step.\\[0.5cm]
{\bf 3.7 Conjecture/Assumption}: We conjecture that in the (chaotic)
regime $QX_0$ correlations have typically an extremely short range due
to the enormous number of links per node and the character of the
local law.\\
Assuming then that the local ''orientations'' of the incident bond
variables are almost statistically independent, both the fluctuations
of the local charge at a typical node $n_i$ and of the charge
difference $(s_i-s_k)$ with respect to neighboring nodes can be
inferred from the '{\it central limit theorem}'(see e.g. \cite{15}),
yielding among other things:
\begin{equation}\Delta s:=<(s-<s>)^2>^{1/2}={\cal
    O}(\Lambda^{1/2})\end{equation}
\begin{equation}\Delta s_{ik}={\cal O}(\Lambda^{1/2})\end{equation}
if we neglect possible phase boundaries (${\cal S}=q\cdot\Ir$). In
other words, local fluctuations are expected to be enormous both with
respect to (clock) time at an arbitrary but fixed node and among nearest
neighbors as $\Lambda$ is assumed to be extremely large.\\[0.5cm]
Remark: Note that the following holds (assuming for simplicity that
$<s_i>=<s_k>$):
\begin{equation}<(s_i-s_k)^2>=<[(s_i-<s_i>)-(s_k-<s_k>)]^2>\end{equation}
which can be written as
\begin{equation}\int[(s_i-<s_i>)-(s_k-<s_k>)]^2\cdot p(s_i)\cdot
  p(s_k)ds_ids_k\end{equation}
with $P(\circ)$ denoting the '{\it probability distribution}' under
discussion, $p(\circ)$ the corresponding '{\it probability
  density}'. For convenience and for the sake of greater generality we
write everything continuously while both our '{\it sample space}' and the
occurring '{\it random variables}' are, by construction,
discrete. (Being sloppy, we identify the random variable, say, $s_i$
with the respective values it can acquire).

As, by assumption, the random variables $s_i,\,s_k$ are independent we
have:
\begin{equation}p(s_i,s_k)=p(s_i)\cdot p(s_k)\end{equation}
and hence get with:
\begin{equation}\int(s_i-<s_i>)p(s_i)ds_i=0\end{equation}
\begin{equation}<(s_i-s_k)^2>=<s_i^2>+<s_k^2>\end{equation}
i.e., up to a trivial factor the same kind of
distribution.\vspace{0.5cm}

The soundness of the above conjecture has to be confirmed by performing
simulations with a variety of local configurations around a given
typical node. Qualitatively one could test this conjecture as follows: Let us e.g. assume
that $s_0$ at node $n_0$ happens to deviate from the surrounding
$s_i's$ in a systematic way at time $t$; to be specific, we assume
$s_0>s_i$ for most of the neighboring nodes. 

This local configuration makes it highly probable that, due tu part B)
of the particular local law introduced above, most of the bonds $b_{0i}$ are ''pointing'' from
$n_0$ to $n_i$ (more precisely: the corresponding $J_{0i}$ will be
positive) after one time step $\tau$, which, after another step
results via part A) in a huge discharge of node $n_0$. As a
consequence the charge $s_0$ happens to be far below the typical
charge of the neighboring sites. This then forces most of the incident
bonds $b_{0i}$ respectively $J_{0i}$ to reorient themselves which
results in a huge surplus charge of node $n_0$ after another two time
steps and so on. Given that all the nodes under discussion are densely
entangled with each other the local dynamics in the phase $QX_0$ may
indeed be sufficiently chaotic to justify our conjecture. What is
however not entirely clear is the statistical weight of such a systematic deviation from
a more or less random distribution of the local charges (see the following
remarks). Furthermore complex systems are capable of performing a lot
of very surprising things (e.g. approaching attractors very quickly
irrespective of their seemingly chaotic behavior; an observation made
by e.g. Kauffmann in his study of '{\it switching nets}, see \cite{17})        \\[0.5cm]
{\bf 3.8 Some Annotations to the Probalisitic Framework}: As to the
soundness of the above conjecture some more remarks are appropriate:\\
i) This is one of the points, mentioned in the introduction, which
comprises in effect a whole bunch of important questions which have a
relevance of their own. The (partly intricate) technical details
coming up in this context shall be postponed for the main part to
forthcoming work (apart from some preliminary remarks following below)
as we consider it to be our main task at the moment to develop in the
rest of the paper a scenario in which space-time is to emerge as a
kind of (coarse grained) order parameter manifold floating in the
discrete network $QX$ and to establish geometric/causal notions like
''nearby'' or ''far away''.\\
ii) Our philosophy is supported by observations made by
e.g. S.Kauffmann and reported in ref. \cite{17}, p.109 or 112,
viz. that too densely connected networks  seem to support chaotic
behavior while more sparsely connected ones are capable of generating
complex behavior. The networks discussed there are however more rigid
than ours in several respects and of a markedly different nature as to
the details of their dynamical laws ('{\it switching nets}').\\
iii) It is of course not really crucial that something like the '{\it
  central limit theorem}' does strictly hold. What we actually do need
is a guarantee that fluctuations tend to be very large and incoherent
and depend to some extent on the density of the wiring. Note that
e.g. for not necessarily independent random varariables $X_i$ (with,
for simplicity, vanishing mean and common variance $\sigma^2$) the
following holds:
\begin{equation}<(\sum_1^n X_i)^2>=\sum_i<X_i^2>+<\sum_{i\neq
    j}X_iX_j>\end{equation}
The first term on the rhs goes as $n\cdot\sigma^2$. If the $X_i$ are
'{\it uniformly weakly correlated}' in the following sense:
\begin{equation}|<X_i\cdot\sum_{j\neq
    i}X_j>|\leq\varepsilon\end{equation}
uniformly in $i$ and $n$ with $\varepsilon<\sigma^2$, we have with
$S_n:=\sum_1^n X_i$:
\begin{equation}n\cdot(\sigma^2-\varepsilon)\leq\,<S_n^2>\,\leq
  n(\sigma^2+\varepsilon)\end{equation}
that is
\begin{equation}\Delta S_n={\cal O}(n^{1/2})\end{equation}
Note that the above assumption is not particularly far fetched as
products like $X_iX_j,i\neq j$, typically oscillate around
zero. Furthermore, similar relations can be established for random
variables which display a certain '{\it cluster behavior}' with
respect to space or/and time (as e.g. in ordinary statistical
mechanics).\\
iv) A careful treatment of many facetts of the central limit theorem
can be found e.g. in \cite{15}. Note that in our case the number of
involved random variables is large but not really infinite. In this
situation the '{\it Berry-Essen-Theorems}' can be applied
(cf. e.g. \cite{19}). Furthermore, in \cite{20} some additional
interesting remarks concerning the possible extension of the central
limit theorem to weakly dependent random variables can be found. Note
in particular that one consequence of the central limit theorem,
viz. the normal distribution of the local fluctuations at, say, a
typical node $n_0$, is not necessarily needed in our scenario,
i.e. the fact that the small fluctuations are the most probable ones
and dominate the behavior.\\
v) Quite the contrary, the particular kind of local fluctuations we
discussed above may rather be an indication for a much more
interesting behavior (not so frequently found in the usual examples of
statistical mechanics), i.e. the tendency of amplification of small
deviations, viz. a kind of '{\it positive feedback behavior}', which
may play perhaps an important role in stabilizing certain phases or, alternatively, drive the system away from perhaps only '{\it metastable}'
phase towards certain '{\it attractors}'. It is however not clear at the moment how frequently these
special fluctuations actually do occur; put differently: how large
their statistical weight is compared to the more chaotic fluctuations
which would rather support a kind of gaussian behavior. This leads to
the last point we want to mention.\\
vi) The above discussion shows that a more elaborated kind of
statistical or stochastic theory is called for if one is dealing with
such peculiar systems (some steps have already been taken in
e.g. refs. \cite{13} and \cite{18} for a however simpler class, the
cellular automata). Note that, in contrast to e.g. Gibbsian
equilibrium statistical mechanics, many pieces of an a priori
framework are missing as a natural probability measure on a suitable
sample space, criteria concerning the relevance and statistical weight
of the various initial configurations, their long-time effects (which
are notoriously difficult to forecast in complex systems) and the
like.

For that reason we made, for the time being, the above heuristic
assumptions and suggest to calculate the various averages and probability
distributions in a more practical way by assuming that the system
behaves reasonable and that the states we are dealing with are
sufficiently generic so that averages and probabilities taken with
respect to e.g. ''space'' and/or ''time'' over, say, one concretely
given actual state of the network (or a time sequence of actual
states) give sensible results (due to the assumption that the huge
numbers of involved nodes and bonds, $|V|,|E|$ may serve as a
substitute for ensemble averages). To give an exemple:\\
vii) The local law shows that fluctuations of the charge $s_0$ at a
given node $n_0$ during one clock time step $\tau$ are given by $\sum
J_{i0}(t)$, the sum extending over the neighboring nodes. If one wants
to make statistical statements about fluctuations at a typical node,
without having an allcomprising statistical framework at ones
disposal, one can proceed as follows:

One chooses e.g. to concentrate on the situation at a fixed clock time
$t$, i.e. make the statistics over the distribution of node and bond
states at a fixed time.\\[0.5cm]
{\bf Definition}: a) The points $\omega$ of the local '{\it sample space}'
$\Omega$ at an abstract typical node $n_0$ are all the possible '{\it
  bond configurations}'
\begin{equation}\{J_{k0}\}_k,\;J_{k0}\in\{+1,0,-1\}\;\mbox{i.e.}\;|\Omega|=3^{deg}\end{equation}
with $deg$ the degree of the node, i.e. the number of incident bonds
$b_{k0}$, $k$ running from $1$ to $deg$.\\
b) Probabilities of '{\it elementary events}' (i.e. a given
configuration) are extracted simply from a frequency analysis over the
array of nodes (frequency of occurrence of the various bond
configurations under discussion) at time $t$ with a suitable but more or less
arbitrary numbering of bonds being implied. More general events can
then be constructed by the usual additivity properties of measures.\\
c) The $J_{k0}(\omega)$ themselves are '{\it elementary random variables}'
with 
\begin{equation}J_{k0}(\omega):=J_{k0}\end{equation}
With the help of the elementary probabilities calculated in b) each of
the random variables $J_{k0}(\omega)$ has a (discrete) distribution
and we can form corresponding sums, i.e:
\begin{equation}J(\omega):=\sum_1^{deg}J_{ko}(\omega)\end{equation}
As charge fluctuations at a node have been linked with the above
$J(\omega)$ which is, on the other side, a sum over elementary random
variables ($deg$ assumed to be very large)
we can make the following observation:\\[0.5cm]
{\bf Observation}: It is $J(\omega)$ and the $J_{k0}(\omega)$ to which
our above assumptions like e.g. the central limit theorem do apply.\vspace{0.5cm}

{\bf 3.9 Corollary to 3.7}: In the perhaps more realistic case ${\cal
  S}=\{0,q,\ldots,q\cdot N\}$ and $N$ large but nevertheless
$\Lambda\gg N$, which we think is natural, given that we assume our
whole universe to emerge from such a $QX_0$, the above estimate shows
that in the regime $QX_0$ the entire local state spaces ${\cal S}_i$
are covered by the expected local fluctuations of $s_i$ with
essentially equal probability $1/N$.\\
In other words, one may say that the '{\it local entropy}' at each
node
\begin{equation}-\sum_{i=0}^N w_i(s_i)\ln w_i(s_i)=-\sum_0^N 1/N\ln
  1/N=\ln N\end{equation}
is maximal in the phase $QX_0$, thus reflecting the absence of any
stable pattern.\\[0.5cm]
Remarks:i) Such (information) entropy concepts may turn out to be quite
useful in analyzing systems like our one (see e.g. \cite{13} or \cite{18}
where related phenomena were analyzed in cellular automata).\\
ii) We would like to note that (vague) resemblances to the scenarios
discussed in, say, '{\it synergetics}' and related fields are not
accidental in our view (cf. e.g. \cite{16} or the beautiful book about
the working philosophy of the Santa Fe Institute; \cite{17} and the
references therein). To various paradigmatic catch words like '{\it
  order parameter}', '{\it slaving principle}' or '{\it selforganized
  criticality}' we hope to come back in forthcoming work.  
\section{The Transition from the Phase $QX_0$ to $QX/ST$ and the
  Emergence of Space-Time as an Order Parameter Manifold}
We provided arguments in the previous section that fluctuations tend
to be extremely large in the primordial phase $QX_0$ and correlations
so short lived that any kind of pattern formation will be obstructed
as long as the fluctuations are on average substantially larger than
the lower critical parameter $\lambda_1$ or, even better, larger than
$\lambda_2$ (version {\bf 2.4 B)} of the local law):
\begin{equation}\Delta
  s_{ik}=<(s_i-s_k)^2>^{1/2}\,>\,\lambda_2\end{equation}
(the average taken with respect to space, time or both).

Let us now assume that, by chance, a sufficiently extended and
pronounced spontaneous fluctuation happens to be created in a '{\it
  subgraph}' $G'(t_0)\subset G$ by the network dynamics around
clock time $t_0$.\\[0.5cm]
{\bf 4.1 Definition}: With $G$ a graph, $V_G$,$E_G$ its sets of
vertices (nodes) and edges (bonds) $G'$ is called a {\it subgraph} if
\begin{equation}V_{G'}\subset V_G\;\mbox{and}\;E_{g'}\subset
  E_G\end{equation}
It is called a {\it section graph} if for every pair of nodes 
\begin{equation}(n_i,n_k)\in
  V_{G'}\,\Rightarrow\,b_{ik}\in E_{G'}\,\mbox{provided
  that}\,b_{ik}\in E_G\end{equation}

We assume this fluctuation to consist of an array of anomalously small
charge differences
\begin{equation}\delta s_{ik}\,\lapprox\,\lambda_1\,\mbox{among the
    nodes of}\,G'\end{equation}
($\lapprox$ meaning that the charge differences are typically smaller
than or approximately equal to $\lambda_1$).\\[0.5cm]
Remark: Note that the subgraph $G'$ need not be connected! Quite the
contrary, it may well be that a subgraph consisting of an array of 
effectively distributed disconnected subclusters will turn out to
serve its purpose much better.\vspace{0.5cm}

If the surrounding network environment is favorable this fluctuation may then trigger an
'{\it avalanche}' of rapidly increasing size in the following way:
\begin{equation}\Delta_s
  s_{ik}\,\lapprox\,\lambda_1\;\mbox{in}\;G'(t_0)\end{equation}
(the index $s$ in $\Delta_s$ denoting a spacial average) will have the effect that a possibly substantial fraction of bonds
$b_{ik}\in E_{G'}$ become temporarily inactive after one clock time
step $\tau$ (and, possibly, for a longer lapse of time ).

By assumption we are still in the chaotic phase $QX_0$. So we expect
that such local temporal deviations from the overall chaotic
''equilibrium state'' will typically quickly dissolve in the rapidly
fluctuating background (at least for the specific network dynamics we
proposed above and for most of the possible scenarios). If, however,
the drop of the average node degree in $G'(t_0+\tau)$ happens to be
pronounced enough so that 
\begin{equation}\Delta_{norm}s_{ik}|_{G'}={\cal
    O}(<deg>_{G'}^{1/2})\lapprox\lambda_2\end{equation}
where the lhs means the generic(!) amount of fluctuations which can be
expected from the assumptions about the environment $QX_0$
(cf. Conjecture/Assumption 3.7), i.e. almost statistical independence
of the orientations of the bonds being incident with the nodes
belonging to $G'$, we may have an entirely new situation!\\[0.5cm]
Remark: $\Delta_ss_{ik}|_{G'(t_0)}$, i.e. the accidental particular
fluctuation which happened to emerge spontaneously at time $t_0$,
should not be confused with the above '{\it typical}' degree of
fluctuations which one has to expect from probability theory. As to
the order of their respective magnitudes we suppose that:
\begin{equation}\Delta_ss_{ik}|_{G'(t_0)}<\Delta_{norm}s_{ik}|_{G'}\end{equation}
even after a certain amount of bonds in $G'$ have already died off in
the course of the phase transition.\\[0.5cm]
{\bf 4.2 Supposition}: We expect that in the course of the phase
transition, which started around clock time $t_0$ the amount of
fluctuations in $G'$ will typically lie below the upper critical
parameter $\lambda_2$ but may on average lie above the lower critical
parameter $\lambda_1$.\vspace{0.5cm}

If the above described scenario actually happens to take place, then
it may occur that the fluctuation pattern prevailing in $G'$
will not be reabsorbed in the background after a few clock time steps
(as has been probably the case many times before) but may represent
the seed for a new and different kind of evolution. Given that the
situation is as being described in Supposition 4.2  there is
a realistic chance that on average more bonds are switched off in and
around $G'$ than are switched on again, i.e:
\begin{equation}N_-(t)-N_+(t)>0\quad\mbox{on average for}\quad t>t_0\end{equation}

It is obvious that the details of this process will depend on the details of the
probability distribution (probability density) for the random variable
$s_{ik}$ restricted to $G'$, this distribution being assumed to be
constructed according to the principles described above. To be more
specific, this kind of evolution will hold under the following
proviso:\\[0.5cm]
{\bf 4.3 Observation}: With $p(s_{ik})|_{G'}$ being the spacial
probability density for the random variable $s_{ik}$ for clock time
$t>t_0$, the unfolding process (i.e. continuing annihilation of bonds)
will go on if
\begin{equation}\int_0^{\lambda_1}p(s_{ik})|_{G'}ds_{ik}>\int_{\lambda_2}^{\infty}p(s_{ik})ds_{ik}\quad\mbox{on
    average}\end{equation}
Remarks: i) As we already mentioned at the end of section 3 such
notions are not entirely easy to define rigorously. Such a probability
distribution should rather be understood in the following way. On the
one side we have a completely deterministic process, a result of which
was (by assumption) this peculiar fluctuation phenomenon at clock time
$t_0$. As it is practically impossible to follow the process in every
detail step by step for $t>t_0$ one has to resort to probabilistic
arguments.

On the basis of a given average node degree $<deg>_{G'}$ and the
assumed statistical independence of bond orientations (in $QX_0$) one
can e.g. calculate (as was shown in section 3) the probability
distribution $p(s_{ik})$ of $s_{ik}$ between two typical neighboring
nodes in $G'$. If this $p(s_{ik})$ fulfills the above inequality for
$t>t_0$ one has reason to expect that the unfolding process will go on
(at least for a while).\\
ii) On the other hand, less stringent assumptions about the stability
of the spontaneous fluctuation at $t_0$ will perhaps already
suffice. What is actually important is that more bonds are annihilated
on average than are created again. This will presumably both depend on
the geometric structure of the subgraph $G'$ and the overall network
state of the graph $G$ around that critical time $t_0$. We will
however postpone a more detailed analysis in this direction and enter
in a description of the new phase $QX/ST$ which we expect to emerge
from this phase transition.\vspace{0.5cm}

In a first step we will simplify our task and concentrate solely on
the geometry of the wiring diagram of the underlying graph $G(t)$, i.e. neglect the
details of the internal states at the nodes and bonds apart from a
bond being dead or alive (as was already done before to some extent),
viz. a bond occurs in the wiring diagram of $G(t)$ if it is alive at
time $t$. This will enable us to extract more clearly the underlying
geometric content being encoded in the network. Furthermore, in order
to limit the amount of technical notation, we will not always
explicitly mention the statistical or fluctuating character of the
various occurring quantities or notions. The correct meaning is
assumed to be tacitly understood in the respective cases without extra
mentioning.

What we are actually after is exhibiting the process in the course of
which what we will later call '{\it physical points}' gradually emerge
from certain protoforms. We assumed that in the primordial phase
$QX_0$ almost all $\Lambda$ nodes have been connected with each other, viz. the
number of bonds in $G_0$ is roughly:
\begin{equation}|E_{G_0}|\approx
  |V_{G_0}|(|V_{G_0}|-1)/2=\Lambda(\Lambda-1)/2\end{equation}
with $\Lambda$ some huge number. Let us now assume that at the onset of
the phase transition a certain (possibly small) fraction of bonds have
become annihilated, say
\begin{equation}1\ll\alpha\ll\Lambda(\Lambda-1)/2\end{equation}
We make, in addition, the idealisation that $G_0$ is (for convenience,
i.e. not only
approximately) a complete graph (simplex). We have then that $\alpha$ arbitrarily selected bonds can at most
connect $k\le 2\alpha$ different nodes, hence there still exist at
least $(\Lambda-k)$ nodes which are maximally connected, viz. they are
spanning a still huge subsimplex $S'\subset G$. On the other hand
there are at most $k\le 2\alpha$ nodes with one or more incident bonds
missing.

This picture will become increasingly intricate if more and more bonds
are switched off. To keep track of a possible emerging '{\it
  superstructure}' we will proceed as follows. We want to cover the
graph $G$ (rather its node set) with a class of particular subgraphs
$\{S_{\nu}\}$, constructed according to the following rule:

Starting from an arbitrary node, say $n_0$, we choose a node $n_1$
being connected with $n_0$ by a bond, and in the i-th step a node being connected with all(!) the preceeding ones; pictorially:
\begin{equation}(n_0\to n_1\to\cdots\to n_k)\end{equation}
The extension process stops if there is no further node $n_{k+1}$
being connectable with all the preceeding ones.\\[0.5cm]
{\bf 4.4 Definitions}: i) A subgraph $S$ is called a '{\it maximal
  subsimplex}'\,($mss$) if there is no other simplex $\hat{S}\neq S$
contained in $G$ with $\hat{S}\supset S$.\\
ii) A subgraph $H$ is called to cover $G$ if $V_H=V_G$ and at every
node $\in V_G$ there exists at least one bond $\in E_H$.\\[0.5cm]
Remark: Such $mss$ are called in combinatorics '{\it cliques}' as we
learned recently.
\vspace{0.5cm}

This class of $mss$, $S_{\nu}\subset G$, and their mutual entanglement
will be a key concept to characterize the geometrical (large scale)
structure of the unfolding graph $G$.\\[0.5cm]
{\bf 4.5 Definition}: Let $G_{\nu}$ be a class of subgraphs of $G$.\\
i) $\cap G_{\nu}$ is the graph with $n\in V_{\cap G_{\nu}}$ if $n\in$
every $V_{G_{\nu}}$,\\
$b_{ik}\in E_{\cap G_{\nu}}$ if $b_{ik}\in$ every $E_{G_{\nu}}$\\
ii) $\cup G_{\nu}$ is the graph with $n\in V_{\cup G_{\nu}}$ if $n\in
V_{G_{\nu}}$ for at least one $\nu$\\
$b_{ik}\in E_{\cup G_{\nu}}$ if $b_{ik}\in E_{G_{\nu}}$ for at least
one $\nu$.\\[0.5cm]
{\bf 4.6 Observations}: i) Starting from an arbitrary node $n_0$ and
performing the above described steps we get a certain $mss$ being
contained in $G$; described pictorially as
\begin{equation}S(n_0\to\cdots\to n_k)\end{equation}
With $S(n_0\to\cdots\to n_k)$ given, each permutation will yield the
same $mss$, i.e:
\begin{equation}S(n_0\to\cdots\to n_k)=S(n_{\pi(0)}\to\cdots\to n_{\pi
    (k)})\end{equation}
Furthermore each $mss$ can so be constructed, starting from one of its
nodes. Evidently this could be done for each node and for all possible
alternatives as to the choice of the next node in the above sequence.

It is however important to note that, starting e.g. from a given node,
there may be several alternatives at each intermediate step, leading
in the end to different(!) $mss$ having some (or even many) nodes and/or bonds in
common! We note in passing that\\
ii) $\cup S_{\nu}$ covers $G$ with $\{S_{\nu}\}$ the class of $mss$. Note
however that in general $\cup S_{\nu}$ is only a true subgraph of $G$,
i.e.
\begin{equation}E_{\cup S_{\nu}}\subset E_G\end{equation}\\[0.5cm]
{\bf 4.7 Example}:\begin{equation}V_G=\{n_0,n_1,n_2,n_3\},\;E_G=\{b_{01},b_{02},b_{12},b_{03},b_{13}\}\end{equation}
There exist two $mss$
\begin{equation}S(n_0\to n_1\to n_2)\;\mbox{and}\;S(n_0\to n_1\to
  n_3)\end{equation}
where at $n_1$ two alternative choices can be made.\vspace{0.5cm}

It turns out to be an ambitious task to calculate the cardinality of
$mss$ in a given graph as a function of the, say, $\alpha$ missing
bonds and their distribution in $G$. the reasons for this are
manifold. For one, this number depends sensitively on the way the
missing bonds are distributed in $G$. For another, as can most easily
be seen by studying examples, the prescription how the $mss$ can be
(re)constructed (cf. Observation 4.6) is strongly ''path dependent''
in the sense that they may be (in particular if they are large)
intricately entangled. which makes an easy counting  quite
delicate. Nevertheless this is an important task.\\[0.5cm]
{\bf 4.8 Task}: Estimate the number of $mss$ spanning $G$ by using only a
few characteristics like the number of missing bonds and the like. It
may well be that also in this situation a statistical approach would
be the most appropriate.\\[0.5cm]
Remark: This problem is presently under study by ourselves. We suppose
that there is some veritable and interesting piece of mathematics
buried in it which goes far beyond the particular context of our
investigation and which would be of relevance in its own right. (Such
problems are typically adressed in \cite{Bollo2} which, we think, will
be of help in this respect). We
conjecture that there are, among other things, structural similarities
to something like (co)homology theory to be built over such discrete
structures.\vspace{0.5cm}

While a general complete solution is at the moment not at our
disposal, we are going to present some preliminary steps and estimates
in that direction. As we already remarked, one might have the idea
that it be possible to provide both a general and sensible upper bound
on the number of $mss$ in a graph $G$ with, say, $\Lambda$ nodes and
$\Lambda(\Lambda-1)/2-\alpha$ bonds. This turns out to be a very
difficult endeavor. It seems to be easier to shift, in order to get a
better feeling for the crucial points of the problem, the point of
view a little bit. That means we want to construct (typical) examples
where the number of $mss$ can be given, thus showing what order of
magnitude one has to expect. 

We construct a special graph as follows: Take $2k$ nodes, choose a
subset $G_1$ consisting of exactly $k$ nodes $(n_1.\ldots,n_k)$, make
$G_1$ a simplex. With the remaining $k$ nodes
$(n_1',\ldots,n_k')$ we proceed in the same way, i.e. we now have two
subsimplices $G_1,G_1'$.

We now choose a one-one-map from $(n_1,\ldots,n_k)$ to
$(n_1',\ldots,n_k')$, say:
\begin{equation}n_i\;\to\;n_i'\end{equation}
We now connect all the $n_i$ with the $n_j'$ except for the $k$ pairs
$(n_i,n_i')$. The graph $G$ so constructed has 
\begin{equation}|E_G|=2k(2k-1)/2-k=2k(2k-2)/2\end{equation}
We see from this that, as in our network scenario, the number of
missing bonds is a relatively small fraction, hence, the example may
be not so untypical.

We can now make the following sequence of observations:\\[0.5cm]
{\bf 4.9 Observations} i) $G_1=S_1$ is already a $mss$ as each
$n_i'\in G_1'$ has one bond missing with respect to $G_1$.\\
ii) One gets new $mss$ by exchanging exactly one $n_i$ with its
partner $n_i'$, pictorially:
\begin{equation}[S_1-n_i+n_i']\end{equation}
yielding $k$ further $mss$.\\
iii) One can proceed by constructing another class of $mss$, now
deleting $(n_i,n_j)$ and adding their respective partners, i.e:
\begin{equation}[S_1-n_i-n_j+n_i'+n_j']\end
{equation}
iv) This can be done until we end up with the $mss$
\begin{equation}[S_1-n_1-\cdots-n_k+n_1'+\cdots+n_k']=S_1'\end{equation}
The combinatorics goes as follows:
\begin{equation}|\{mss\}|=\sum_{\nu=0}^k {k \choose
    \nu}=(1+1)^k=2^k\end{equation}
i.e., our $2k-$node-graph (with $k$ bonds missing) contains exactly
$2^k$ $mss$.\\[0.5cm]
{\bf 4.10 Some Comments on Combinatorics}: That the combinatorical
problems we have adressed above are really ambitious can be seen with
the help of the following argument. As a ''warm-up exercise'' one can
tackle a seemingly much simpler problem, namely trying to find
sensible bounds for the cardinality of certain coverings $C\in{\cal
  C}$ of a given finite (but, in general, large) set $X$ with $C$
being a covering of $X$, $C\in P(X)$, the '{\it power set}' of $X$. The constraint is that in a given covering $C$ no subsets of
already occurring sets are to be allowed, i.e., 
\beq X_i\in C\;\mbox{implies}\;C\ni X_j\not\subset X_i\;X_i\not\subset
X_j\;\mbox{for}\;j\neq
i\eeq
The reason for this particular constraint stems from our $mss$ (see
Corollary 4.12 below).

We are e.g. interested in bounds of the following nature:\\[0.5cm] 
Give an
effective bound for $sup(|C|;\,C\in {\cal C})$, i.e. a bound which is
better than $|P(X)|$, by employing the above constraint.\\[0.5cm]
As for our graph problem discussed above, this appears to be still
quite difficult if one starts from first principles. Shifting the point of view one can instead try to
construct coverings $C$ with  a large $|C|$ in order to get at least a
sensible lower bound on $sup(|C|)$. It turns out to be effective to
choose coverings with sets $X_j$ with $|X_i|=|X_j|$ for all $j$ and
then take $|X_i|=k/2$ if $|X|=k$ with $k$ being even. (This idea we owe
to our coworker Th. Nowotny; note that in that case the above
constraint is automatically implied). Then we have ($k$
large):
\beq |C|=k!/([k/2]!)^2\approx (2/\pi k)^{1/2}\cdot 2^k\eeq
with the help of Stirling's formula. Note that a covering with
non-overlapping sets (i.e. a '{\it partition}') has always a
cardinality $|C|\leq k$.

A little bit surprisingly, this construction happens already to be
optimal, as we found out recently when we stumbled by chance over
an old but noteworthy paper by E.Sperner, \cite{sperner} (see also the beautiful book
of H.Lueneburg, \cite{luene} chapt.XVI and the remark on p.492 or \cite{Bollo1}). In
this paper the following result has been proven (which, by the way
gave rise to a full-fledged subtheory in combinatorics);
(in our notation):\\[0.5cm]
{\bf Sperner's Theorem}: 
\beq sup\,|C|={n \choose k}\;\mbox{with}\;k=[n/2]\eeq
and $sup$ is attained for $P_k(X)$, a covering with $k-$sets, if $n$
is even, $P_k(X)$ and $P_{k+1}(X)$ if $n$ is odd.\\[0.5cm]
Remark: It may well be that the methods developed there may be also
helpful in our more general graph-problem.\vspace{0.5cm}

Returning to our concrete network scenario we can now proceed as
follows:\\
$V_G$ can be split in the following way:
\beq V_G=V_{S'}\cup V_N\eeq
with $V_N$ the set of nodes with some of the $\alpha$ bonds among them
missing, $V_{S'}$ the set of nodes being maximally connected (see the
remarks after formula (53))
\beq |V_N|=k\leq 2\alpha\eeq
Almost by definition, $V_{S'}$ generate a simplex $S'\subset
G$.\\[0.5cm]
{\bf 4.11 Observation}: i) The simplex $S'$ is contained in each of
the maximal subsimplices $S_{\nu}$, i.e:
\beq S'\subset\cap S_{\nu}\;\mbox{with, in general,}\;S'\neq\cap
S_{\nu}\eeq
ii) $S'$ itself is not(!) maximal as $[S'\cup n_i]$ is always a larger
simplex with $n_i\in N$ and $[S'\cup n_i]$ being the section graph
spanned by $V_{S'}$ and $n_i$.\\
iii) To each maximal simplex $S_{\nu}\subset G$ belongs a unique
maximal subsimplex $N_{\nu}\subset N$ with
\beq S_{\nu}=[S'\cup N_{\nu}]\eeq \\[0.5cm]
{\bf 4.12 Corollary}: From the maximality of the $N_{\nu}$ follows a
general structure relation for the $\{S_{\nu}\}$ and $\{N_{\nu}\}$:
\beq \nu\neq\mu\,\to\,S_{\nu}\neq S_{\mu}\,\to\,N_{\nu}\neq
N_{\mu}\eeq
and neither
\beq N_{\nu}\subset N_{\mu}\;\mbox{nor}\;N_{\mu}\subset N_{\nu}\eeq
viz. there always exists at least one $n_{\nu}\in V_{N_{\nu}}$
s.t. $n_{\nu}\notin V_{N_{\mu}}$ and vice versa.\\[0.5cm]
Proof of Observation 4.11: i) Starting from an arbitrary node $n\in
G$, it is by definition connected with all the other nodes in $S'$,
since if say $n,\,n'$ are not connected they both belong to $N$ (by
definition). I.e., irrespectively how we will proceed in the
construction of some $S_{\nu}$, $S'$ can always be added at any
intermediate step, hence $S'\subset\cap S_{\nu}$. On the other side
one can easily construct scenarios where $S'\neq\cap S_{\nu}$.\\
ii) As $n\in N$ is connected with each $n'\in S'$ (by definition of
$N$ and $S'$), the section graph $[S'\cup n]$ is again a (larger)
simplex.\\
iii) We have $S'\subset S_{\nu}$ for all $\nu$, hence 
\beq S_{\nu}\neq S_{\mu}\;\mbox{implies}\;N_{\nu}\neq N_{\mu}\eeq
with $N_{\nu,\mu}$ the corresponding section graphs in $N$.\\
With $S_{\nu}$ being a simplex, $N_{\nu}$ is again a subsimplex which
is maximal in $N$. Otherwise $S_{\nu}$ would not be maximal in $G$.\\
On the other side each $S_{\nu}=[S'\cup N_{\nu}]$ is uniquely given by
a maximal $N_{\nu}$ in $N$ as each node in $N$ is connected with all
the nodes in $S'$.\vspace{0.5cm}

We see from the above that as long as $\alpha$, the number of dead
(missing) bonds, is much smaller than the number of bonds in the
initial simplex $G_0$, there does exist a considerable overlap $S'$, among the class of $mss$ $S_{\nu}$. This
overlap will become smaller with $\alpha$ increasing with clock time
$t$; by the same token the number of $mss$ will increase. To describe
this unfolding process we make the following pictorial abbreviations
and observations:\\[0.5cm]
{\bf 4.13 Abbreviation}: We abbreviate $n_i,\,n_k$ (not) connected by
a bond by
\beq n_i\sim n_k\quad (n_i\not\sim n_k)\eeq\\[0.5cm]
We then have:\\[0.5cm]
{\bf 4.14 Observation}: i) $n_i\not\sim n_k$ implies that they are
lying in different $S_{\nu}$'s.\\
ii) $S_{\nu}\,,\,S_{\mu}$ are disjoint, i.e. $S_{\nu}\cap
S_{\mu}=\emptyset$ iff 
\beq \forall n_{\nu}\in S_{\nu}\;\exists\; n_{\mu}\in
S_{\mu}\;\mbox{with}\;n_{\nu} \not\sim n_{\mu}\eeq
or vice versa.\\[0.5cm]
{\bf 4.15 Consequence}: This shows that it may well be that
$S_{\nu}\cap S_{\mu}=\emptyset$ while the two $mss$ have still a lot of
'{\it interbonds}', i.e. bonds connecting the one with the other. The
guiding idea is however that $V_{S_{\nu}}$ and $V_{S_{\mu}}$, taken as
a whole, will be generically considerably less strongly entangled with
each other than the nodes within $S_{\nu}$ or $S_{\mu}$ among
themselves after the unfolding process is fully developed.\\[0.5cm]
{\bf 4.16 The Physical Picture}: For $\alpha$ small, viz. $S'\subset
G$ still large, we regard the emerging $S_{\nu}$'s as '{\it
  protoforms}' of '{\it physical points}'. We suppose that
this picture becomes more pronounced with increasing $\alpha$,
i.e. increasing clock time, when the entanglement between different
proto points becomes weaker. For $\alpha$ sufficiently small all these
proto points are hanging together via the non-empty(!) $S'$ while for
$\alpha$ sufficiently large (i.e. clock time large) and the dead bonds
appropriately distributed over $G$ it may happen that $S'=\emptyset$
and that, furthermore, a pronounced far- and near-order among the
grains $S_{\nu}$ is established via their (varying) '{\it degree of
  connectedness}'.

This is the scenario which, we hope, will support a certain '{\it
  superstructure}' we dub $ST$ (space-time), the whole complex we like
  to call $QX/ST$, i.e. a still wildly fluctuating ''quantum
  underworld'' with a both coarser and more smoothly behaving '{\it
  order parameter manifold}' being superimposed.\vspace{0.5cm}

One can supply a rough estimate as to the threshold which (under
certain conditions) may divide these two scenarios. With the
primordial graph $G_0$ being nearly a simplex of node number, say,
$\Lambda$, edge number $\Lambda(\Lambda-1)/2$ and $\alpha$ bonds
already annihilated, we had the result (see above) that $S'$ consists
at least of
\beq |V_{S'}|\geq\Lambda-2\alpha\eeq
nodes as long as the rhs is positive. Hence:\\[0.5cm]
{\bf 4.17 Observation}: There is a chance that $S'=\emptyset$, i.e an
effective distribution of the annihilated bonds being assumed, when
\beq \alpha\geq\Lambda/2\eeq
We conjecture that this is a characteristic number which indicates
where the transition zone will probably lie which divides the two
regimes described above. The threshold value yields:
\beq \Lambda(\Lambda-1)/2-\Lambda/2=\Lambda(\Lambda-2)/2\eeq
bonds being still alive, which implies an average node degree:
\beq <deg>_s=\Lambda-2\eeq
in other words, on the average there is only one bond missing per
node!\\[0.5cm]
Remark: The content of Observation 4.17 makes it clear again that
relatively advanced stochastic graph concepts are called for, of a
kind as they are developed in the beautiful book of Bollobas about
'{\it random graphs}' (\cite{Bollo2}), i.e. it has to be clarified
whether e.g. $\alpha=\Lambda/2$ is actually representing what is
called a '{\it threshold function}' for random graphs. These extremely
important concepts will be discussed in more detail elsewhere, as we
want to enter in the remaining part of this paper into a more thorough
analysis of the nature of '{\it physical points}' we have introduced
above and their mutual entanglement, in particular for large $\alpha$
i.e. far away from the phase transition regime.\vspace{0.5cm}

Our strategy to relate the $mss$ $S_{\nu}$ with what one may call on a
much more coarse grained level physical points suggests the
introduction of the following mathematical concepts, being of wide use
in combinatorial mathematics and related fields
(cf. e.g. \cite{Beutel} or \cite{Halder}).\\[0.5cm]
{\bf 4.18 Definition}: Over the '{\it ground set}' $V$ we construct an
'{\it incidence structure}' $(V,{\cal S},I)$ or '{\it block space}' in
the following way:\\
i) The '{\it blocks}' are the $mss$ $\{S_{\nu}\}$\\
ii) The points of the incidence structure are the nodes\\
iii) The '{\it incidence relation}' $I$ is $\in$.\\
In other words we say a node is incident with a block $S_{\nu}$ or
belongs to $S_{\nu}$. In our scenario the blocks are elements of the
power set $P(V)$ (which needs not always be the case).\\
iv) With $(n)$ we denote the blocks $S_{\nu}$ which are incident with
$n$, i.e. $n\in S_{\nu}$. In the general theory $(S_{\nu})$ would
denote the nodes being incident with $S_{\nu}$, in our case where
$I\doteq\in$ we identify $(S_{\nu})$ and $S_{\nu}$;
$|(n)|,\,|S_{\nu}|$ are their respective cardinalities.\\
v) The above is generalized in an obvious way to
$(n_{i_1},\ldots,n_{i_k})$ or $(S_{\nu_1},\ldots,S_{\nu_k})$.\\[0.5cm]
{\bf 4.19 Definition}: i) An '{\it automorphism}' $A$ of a block space
is a bijective map $V\to V$ s.t.
\beq A:\;S_{\nu}\to S_{nu'}\,,\;\{\nu'\}\,\mbox{a permutation
  of}\,\{\nu\}\eeq
ii) The automorphism $A$ is called '{\it inner}' if it acts within the
respective blocks, i.e:
\beq A:\;S_{\nu}\to S_{\nu}\quad\mbox{for all}\,\nu\eeq\\[0.5cm]
{\bf 4.20 Definition}: With respect to the above block space we can
speak of an\\
i) '{\it interior bond}' of a given $S_{\nu}$, i.e:
\beq b_{ik}\;\mbox{with}\,n_i,\,n_k\,\in S_{\nu}\eeq
ii) '{\it exterior bond}' with respect to a given $S_{\nu}$, i.e:
\beq b_{ik}\;\mbox{with}\,n_i,\,n_k\,\notin S_{\nu}\eeq
iii) an '{\it interbond}', i.e:
\beq b_{ik}\;\mbox{with}\,n_i\in S_{\nu},\,n_k\in
S_{\mu},\,\nu\neq\mu\eeq
iv) a '{\it common bond}' of $S_{\nu}$,$S_{\mu}$ if $b_{ik}$ is an
interior bond both of $S_{\nu}$ and $S_{\mu}$.\\
v) a '{\it true interbond}' $b_{ik}$ if for $\nu\neq\mu$:
\beq n_i\in S_{\nu},\,n_k\in S_{\mu},\,n_k\notin S_{\nu}\eeq
vi) We then have the relation for given $S_{\nu},\,S_{\mu}$:
\beq \{interbonds\}-\{common\;bonds\}=\{true\;interbonds\}\eeq
Remarks: i) The relation between these classes describes the (time
dependent) degree of entanglement among the blocks $S_{\nu}$,
viz. among the physical proto points and, as a consequence, the
physical near- and far-order on the level of $ST$, the macroscopic
causality structure of space-time and the non-local entanglement we
observe in quantum mechanics.\\
ii) In Definition 2.1 we defined the notion of a '{\it simple graph}' and
related it with the concept of a '{\it homogeneous}', '{\it non-reflexive}' 
relation. Note that, on the other side, a graph is a particular
example of a block space, with the class of bonds being the blocks and
the obvious incidence relation between nodes and bonds.

This identification generalizes easily to more complex scenarios in
the following way: Following Bollobas in \cite{Bollo1} we define a
'{\it set system}' ${\cal F}$ over, say, $V$ to be a subset of the
powerset $P(V)$, i.e:
\beq {\cal F}\subset P(V)\eeq
Such a set system is by the same token a block space, both being, on the
other side, examples of a '{\it heterogeneous relation}' between $V$
and $P(V)$ ( a homogeneous relation corresponding to blocks
$\subset P^{(2)}(V)$).\\[0.5cm]
{\bf 4.21 Definition}: i) $P^{(k)}(V)$ is the class of $k-sets$ in
$P(V)$\\
ii) A homogeneous relation is a subset of $V\times V$ or, by the same
token, a subset of $P^{(2)}(V)$\\
iii) A heterogeneous relation is a relation between between two sets
of, typically, different nature, i.e. a block space. If the set of
blocks is contained in $P(V)$ one may also call it a '{\it hyper
  graph}' and its blocks, i.e. the occurring tuples of nodes its '{\it
  hyper edges}'. If the hyper edges are uniformly taken from
$P^{(k)}(V)$ it is called a '{\it k-uniform}' hyper graph (see
e.g. \cite{Bollo1})\\
iv) As is the case for simple graphs and block spaces, one can now
talk of the '{\it degree of a node}' in a hypergraph, i.e. the number
of incident hyper edges, i.e. $|(n)|$ or the rank of a hyper edge,
which is in our example simply the number of nodes belonging to
it.\\[0.5cm]
Remark: We would like to emphasize that, while this various
definitions and introduced concepts seem to be, at first glance, more
or less straightforward and related, they nevertheless turn out to be
very efficient and economical tools to deal with a number of
surprisingly deep questions in discrete mathematics from different
perspectives (e.g. graphs, relational mathematics, incidence
structures or block spaces; see the above mentioned
literature).\\[0.5cm]
{\bf 4.22 Observation}: As a simplex is uniquely given by its set of nodes,
i.e. a set $\subset P(V)$, we can associate our complex $QX/ST$,
i.e. the underlying graph $G$ and the array of $mss$, with a
hypergraph by identifying the $S_{\nu}$'s with its hyperbonds. This
hypergraph is nothing but $\cup S_{\nu}\subset G$ (introduced in
Observation 4.6), viz. it exhibits in general not the full wiring of
$G$ or $QX$ but only the bonds occurring in the $S_{\nu}$'s, the true
interbonds, however, are missing in $\cup S_{\nu}$.\vspace{0.5cm}

In a next step one can go to a coarser level of resolution by defining
the '{\it intersection graph}' of the set $\{S_{\nu}\}$; henceforth
(as the notion hypergraph is already fixed in graph theory in the
sense defined above) we prefer to call it the associated '{\it skeleton
  graph}' or '{\it super graph}'.\\[0.5cm]
{\bf 4.23 Observation}: By shrinking the blocks $S_{\nu}$ to '{\it
  super nodes}' one gets another graph by saying that two super nodes
$S_{\nu}$, $S_{\mu}$ ($\nu\neq\mu$), are linked by a '{\it super
  bond}' if their intersection $S_{\nu}\cap
S_{\mu}\neq\emptyset$. This '{\it super graph}' may then be associated
with the manifold of '{\it unresolved}' physical points, $ST$, if we
neglect their internal complexity (cf. this with what we said in 4.15,
4.16 above).\vspace{0.5cm}

The above train of thought shows that one may impose on the underlying
network $QX$ a certain hierarchy of levels of varying resolution of
the physical landscape which becomes more and more pronounced with
increasing number of turned-off bonds, viz. with increasing clock
time. Given an arbitrary but fixed block $S_0$, one can define its
infinitesimal neighborhood $U_1(S_0)$ as consisting of the $mss$ $S^0_{\nu}$ with
\beq S^0_{\nu}\cap S_0\neq\emptyset\eeq
in other words, the nodes adjacent to $S_0$ in the super graph $ST$ of
4.23.

There may then exist $mss$ $S_{\mu}$ with
\beq S_{\mu}\cap S^0_{\nu}\neq\emptyset\;\mbox{for
  some}\,\nu\;\mbox{but}\;S_{\mu}\cap S_0=\emptyset\eeq
i.e. super nodes directly connected with some $S^0_{\nu}$ but no
longer with $S_0$ itself, viz. they belong to $U_1(S_{\nu})$ but only
to the second order neighborhood $U_2(S_0)$ of $S_0$ (as for this
topological concept cf. Definition 2.1 above). Proceeding in this way
it may be possible to impose some metrical near- and far-order on the
network by taking in a certain approximation only the information into
account which is encoded in the mutual overlap of the $mss$ $S:{\nu}$
i.e. in $ST=\cup S_{\nu}$. What is projected out on this level are the
(possibly still numerous) interbonds which may exist between the
various disjunct $mss$. The details of the neighborhood structure
encoded in the full $QX/ST$ can then be rebuilt starting from the
skeleton graph $ST=\cup S_{\nu}$.

In the physics of many degrees of freedom what typically matters is
the strength of interaction between the various
constituents. Furthermore this characteristic quantity is usually
closely related with their spatial distance in the system under
discussion and its dimension. Following these lines we will also
organize our network.

Given two node sets $A,\,B$ or the respective subgraphs we can count
the number of bonds between them and regard this as a measure of their
mutual dynamical coupling.\\[0.5cm]
{\bf 4.24 Definition (Connectivity)}: With $A,\,B$ being two sets of
the above kind we denote by$|A\sim B|$ the number of bonds connecting
the nodes of $A$ with the nodes of $B$ and by $|A\sim B|_m$ their
maximal possible number. Then we call
\beq 0\leq c_{AB}:=|A\sim B|/|a\sim B|_m\leq 1\eeq
the '{\it connectivity}' of the pair $A,\,B$.\\
It represents the probability that a randomly picked up pair of nodes
$n_A\in A,\,n_B\in B$ is connected by a bond.\\
$|A\sim B|_m$ depends however on the mutual relation between $A$ and
$B$!\\[0.5cm]
{\bf 4.25 Observation}: i) 
\beq A\cap B=\emptyset \to |A\sim B|_m=|A|\cdot|B|\eeq
($|A|,\,|B|$ the respective number of nodes), hence
\beq c_{AB}=|A\sim B|/|A|\cdot|B|\eeq
ii)
\beq A=B\to |A\sim B|_m={|A| \choose 2}\eeq
iii)
\begin{eqnarray} 
A\cap B\neq\emptyset\to|A\sim B|_m & = & |(A-B)\sim(B-A)|_m\\
                                   &   & +|(A\Delta B)\sim(A\cap
                                   B)|_m+|(A\cap B)\sim(A\cap
                                   B)|_m\nonumber
\end{eqnarray}
i.e:
\beq |A\sim B|_m=|A-B|\cdot|B-A|+|A\cap B\cdot(|A-B|+|B-A|)+{|A\cap B|
  \choose 2}\eeq
with $A\Delta B$ being the symmetric difference of $A$ and
$B$.\\[0.5cm]
Remark: There exist of course other possibilities to quantify the
degree of mutual influence among the various regions of the graph. One
could e.g. admit not only direct bonds as connecting paths but paths
up to a certain length etc. Furthermore there are a variety of (not
primarily physically motivated) notions of connectedness in use in
graph theory (e.g. the theorems of Menger, cf. \cite{5}). The concept
we developed above is adapted to our particular scenario with the
$mss$ as building blocks but with possibly a lot of surviving
interbonds between disjunct grains $S_{\nu},?,S_{\mu}$ which may
perhaps be quite a distance apart on the skeleton graph $ST$. Our
notion is (among other things) able to measure these residual bonds
which are in our view responsible for the observed '{\it
  non-locality}' of quantum theory.\\[0.5cm]
{\bf 4.26 The Metric/Topological Picture of $QX/ST$ (a first Draft)}:\\
The picture we expect to emerge after the unfolding process is fully
developed (i.e. $\alpha$ relatively large) is now the following:\\
i) It is one of the many fascinating observations made in
\cite{Bollo2} that, perhaps against the usual intuition, random graphs
tend to be surprisingly regular, i.e. the generic(!) graph with, say, $n$
nodes and $m$ bonds tends to be almost `{\it translation invariant}`,
viz. the node degree is roughly the same over the graph (an effect of
the peaked probability in phase space, similar to e.g. related
phenomena in statistical mechanics). In other words, it is perhaps not
too far-fetched to expect the same, at least on average, for the
typical shape of the $mss$ and their entanglement.\\
ii) Taking then a typical grain $S_0$ of $ST$, we expect its
infinitesimal neighborhood $\{S^0_{\nu}\}$ to be densely connected
with $S_0$ in the sense of Definition 4.24. In other words:
\beq c_{S_0S_0}=1\quad,\quad c_{S_0S^0_{\nu}}\lapprox 1\eeq
iii) Proceeding further with picking up grains $S_{\mu}$ in decreasing
order of $c_{S_0S_{\mu}}$ we can construct shell after shell around
$S_0$ with weaker and weaker connectivity as regards to the central
element $S_0$, i.e:
\beq 1\geq c_{S_0S^0_{\nu}}\geq c_{S_0S^1_{\nu'}}\cdots\eeq
iv) We expect this process to be consistent with the neighborhood
structure on the super graph $ST$ which is defined by intersection,
i.e. that node distance on $ST$ corresponds more or less with the
decrease in connectivity (properties like these have of course to be
checked more systematically, analytically and/or via simulations on a
computer which is under way).
\section{Comments and Concluding Remarks}
In this last section we would like to briefly recapitulate what has
been accomplished in the preceeding sections and what remains to be
done in forthcoming work. Furthermore we will focus on some
longstanding goals which, we hope, can be acchieved by continuing the
train of ideas developed and sketched above.

We have introduced and studied a class of cellular network models
(rather, some simple candidates of a presumably much larger set) the
dynamical laws of which are capable of performing some peculiar sort
of {\it unfolding} in addition to the ordinary dynamics. This is
implemented by introducing a socalled {\it hystheresis interval} given
by the upper and lower critical parameters $\lambda_2,\;\lambda_1$
which regulate the switching-on and -off of elementary interactions
between the nodes of the network.

We argue that the network evolution will drive the system for generic
initial conditions  towards a special type of attractor which, we
conjecture, is the underlying hidden discrete substratum of our
continuous space-time manifold (or rather, the physical vacuum) as we
know it on a much coarser scale of resolution.

As particular constituents we describe the formation of physical
(proto)points which consist in our scheme of densly entangled
subclusters of, as we expect, roughly Planck scale size. The complex
internal structure of these subclusters ($mss$) is supposed to be the
carrier of the (quantum)fields which occur in continuum physics as
elementary building blocks. This transition from the discrete network
substratum to the continuum description shall be worked out in greater
detail in forthcoming work together with a {\it correspondence
  principle} which relates both gravity and quantum theory with
certain aspects of our scenario.

As to the technical side of the investigation, simulations and
implementations of the various laws together with an appropriate
sample of choices of the respective parameters and initial conditions
are presently performed on a computer. They are designed to test
and/or verify the conjectured scenarios described above. In this
connection many details can and have to be checked like {\it length of
  transients, type of attractors, number, size and internal structure
  of the $mss$ and their mutual overlap, cycle length, average node degree, number of extinct
  bonds, strength of fluctuations} and the like. The results shall be
published in the near future.

Last but not least, we would like to relate our approach in more
detail with the ideas of e.g. Sorkin et al., the conceptual scheme
provided by Isham and the train of ideas developed by 't Hooft (see
the references), not to mention the whole field of non-commutative
geometry as such (see however our first paper \cite{1}). We regret
that this could not be done already in this paper due to lack of
space. Furthermore, the deep results of the fascinating field of {\it random graphs} are only very briefly touched.\\[2cm]
Acknowledgement: We would like to thank H.L.de Vries for many helpful
remarks concerning the existing literature about various fields of
discrete mathematics.

\end{document}